\documentclass{iopconfser}
\usepackage{graphicx}
\usepackage{dcolumn}
\usepackage{bm}
\usepackage{amsmath}
\usepackage{color}

\begin{document}

\title{Effect of magnetic field on whirling-anti-whirling order in icosahedral-quasicrystal approximant}

\author{Shinji Watanabe, Tatsuya Iwasaki}

\affil{Department of Basic Sciences, Kyushu Institute of Technology, Kitakyushu, Fukuoka 804-8550, Japan}

\email{swata—mns.kyutech.ac.jp}

\begin{abstract}
Recent neutron measurement in the icosahedral quasicrystal approximant Au-SM-Tb (SM=Al, Ga) has revealed unique noncollinear magnetic order ``whirling-anti-whirling states''. 
Here, we report theoretical analysis on the magnetic-field-direction dependence on the whirling-anti-whirling order in the 1/1 approximant crystal. By performing exact-diagonalization calculation for the effective model taking into account the uniaxial magnetic anisotropy arising from the crystalline electric field, we show the metamagnetic transition takes place simultaneously with the topological transition under the magnetic field along the (111) direction. After the metamagnetic transition, the emergent fictious magnetic field induced by the chirality of noncoplanar magnetic moments appears, the analysis of which concludes that the topological Hall effect is expected to be observed in the electrical conductivity $\sigma_{xy}$ and $\sigma_{yz}$ for the applied field direction from (111) to (001). 
\end{abstract}

\section{Introduction}

Quasicrystal (QC) has no periodicity of lattice but possesses unique structural order, which is distinct from amorphous. 
Since the discovery by D. Schechtman in 1984~\cite{Shechtman} the understanding of the QC structure of atomic arrangement has proceeded~\cite{Tsai,Takakura}. 
However, the electronic state and physical property are far from complete understanding because the Bloch theorem can no longer be applied. 

\begin{figure}
\begin{center}
\includegraphics[width=14cm]{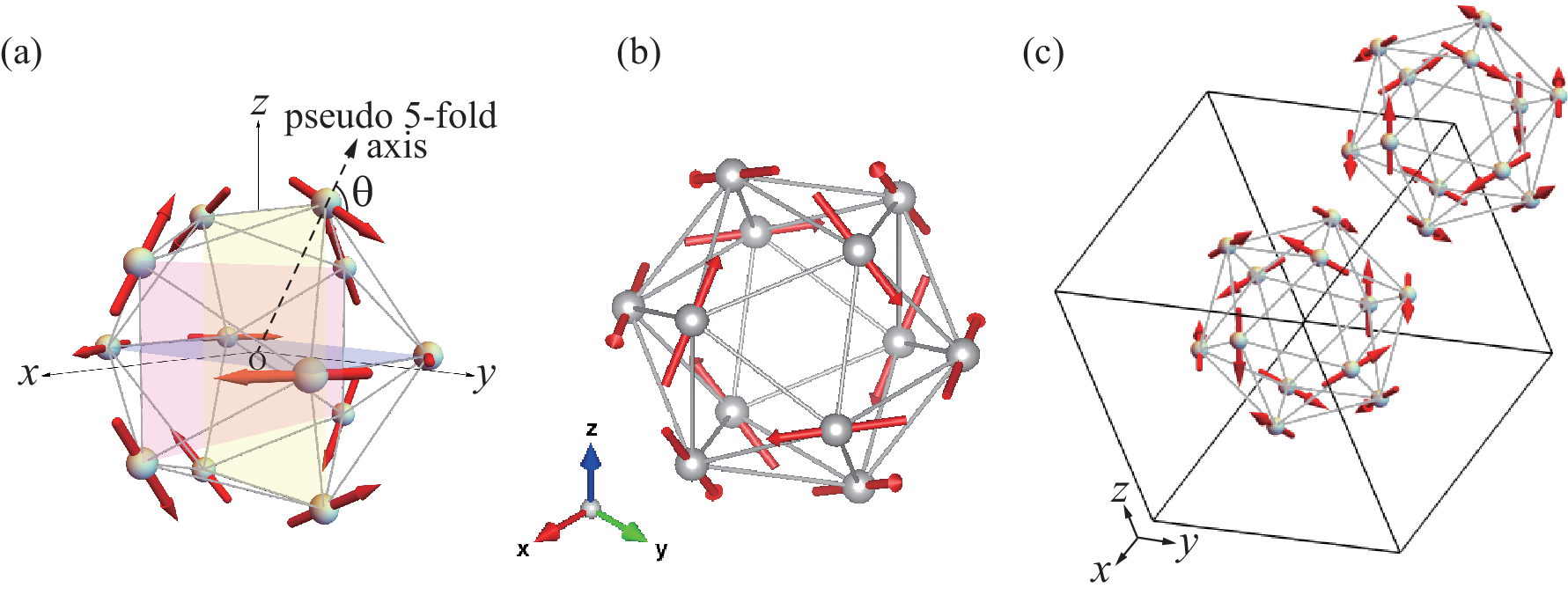}
\caption{
(a) Whirling-moment state on the IC for $\theta=90^{\circ}$. The vector passing through the rare-earth site from the center of the IC is the pseudo 5-fold axis drawn with the dashed line with an arrow indicates the pseudo 5-fold axis, which is in the mirror plane colored by yellow surface. The same is applied to each rare-earth site where the mirror planes are colored by pink and purple surfaces. 
(b) Whirling moment state on the IC seen from the (111) direction. 
(c) Whirling-moment state at the center IC and the anti-whirling-moment state at the corner IC in the 1/1 AC viewed from the (111) direction. The frame box is the bcc unit cell.
}
\label{fig:W_AW}
\end{center}
\end{figure}

Recently, the rare-earth based icosahedral QCs have attracted great interest since the ferromagnetic (FM) long-range order has been discovered experimentally in the Au-Ga-R (R=Tb, Gd, and Dy)~\cite{Tamura2021,Takeuchi2023}. 
There also exists approximant crystal (AC), which has the local atomic configuration common to that in the QC with periodicity. 
In the 1/1 AC Cd$_6$R (R=Pr, Nd, Sm, Gd, Tb, Dy, Ho, Er, Tm, and Lu)~\cite{Tamura2010,Mori,Tamura2012} and 1/1 AC Au-SM-R (SM=Si, Ge, and Al; R=Gd, Tb, Dy, and Ho)~\cite{Hiroto2013,Hiroto2014,Das,Sato2019,Hiroto,Gebresenbut,Nawa2023,Labib2024} as well as the 2/1 AC~\cite{Inagaki,So}, the FM long-range order and the antiferromagnetic (AFM) long-range order have been observed~\cite{Suzuki}. 
These QCs and ACs are composed of the Tsai-type cluster containing the icosahedron (IC), at the 12 vertices of which the rare-earth atoms are located as shown in Fig.~\ref{fig:W_AW}(a). 
In the 1/1 AC, the IC is located at the center and corner of the body-center-cubic (bcc) lattice.  
The 4f electron at the rare-earth site is responsible for the magnetism in the QCs and ACs. 

Recently, the theory of crystalline electric field (CEF) in the QC and AC has been formulated on the basis of the point charge model~\cite{WK2021}. 
The analysis of the CEF in the Tb-based AC has revealed that the magnetic easy axis of the CEF ground state is lying in the mirror plane located at each Tb site~\cite{WPNAS}. 
The analysis of the CEF in the Dy-based AC has also shown that the magnetic anisotropy is similar to that in the Tb-based systems~\cite{WI2023}. 
In Fig.~\ref{fig:W_AW}(a), the mirror planes are illustrated as colored planes of yellow, pink, and purple. 
The vector passing through each Tb site from the center of the IC drawn with the dashed line with an arrow indicates the pseudo 5-fold axis. 
In Refs.~\cite{WPNAS,WSR}, the angle $\theta$ between the magnetic easy axis and the pseudo 5-fold axis in the mirror plane was defined as Fig.~\ref{fig:W_AW}(a) and the ground state in the magnetic model on the IC was identified for various anisotropy angle $\theta$ by numerical calculations.

Interestingly, depending on the anisotropy angle $\theta$, the various magnetic states appear, among which non-trivial topological magnetic textures were discovered theoretically~\cite{WPNAS,WSR}. 
One of them is the whirling-moment state, whose magnetic structure with $\theta=90^{\circ}$ is illustrated in Fig.~\ref{fig:W_AW}(a). This state has the swirling magnetic moments on the IC, which is clearly seen from the (111) direction as shown in Fig.~\ref{fig:W_AW}(b). 

In Refs.~\cite{WPNAS,WSR}, the ground-state phase diagram of the 1/1 AC was theoretically analyzed on the basis of the magnetic ground states obtained for the single IC, where the whirling-anti-whirling order is realized as shown in Fig.~\ref{fig:W_AW}(c). 
Here, the anti-whirling-moment state is the state with all the magnetic moments being inverted from those in the whirling-moment state, which is located at the corner IC in the bcc unit cell in Fig.~\ref{fig:W_AW}(c) while the whirling-moment state is located at the center IC.  
Recently, theoretical study on the ground state in the 1/1 AC without assuming the magnetic state on the single IC has been performed and the whirling-anti-whirling state has been confirmed to be realized around $\theta=90^{\circ}$~\cite{WI2024}, which is consistent with the previous study~\cite{WPNAS}.  
Experimentally, the whirling-anti-whirling orders with the anisotropy angles around $\theta\approx 90^{\circ}$ were identified by the neutron measurement in the 1/1 AC Au-Al-Tb~\cite{Sato2019} and Au-Ga-Tb~\cite{Nawa2023,Labib2024}. The magnetic space group was identified experimentally as $I_{p}m'\bar{3}'$~\cite{Labib2024}.

In this study, to get insight into the effect of the magnetic field on the whirling-anti-whirling state, we analyze the ground-state property under the magnetic field. 
By performing the exact-diagonalization calculation, we show that the metamagnetic transition takes place simultaneously with the topological transition under the magnetic field along the (111) direction. 
We analyze the applied field-direction dependence of the emergent fictious magnetic field, which predicts that the topological Hall effect appears in the electrical conductivity $\sigma_{xy}$ and $\sigma_{yz}$ for the applied field direction from (111) to (001).

\section{Model and method}

We consider the effective model~\cite{Sato2019,WPNAS,WSR,WI2024,WI2024b}
\begin{eqnarray}
H=-\sum_{\langle i,j\rangle}J_{ij}\hat{\bm J}_i\cdot\hat{\bm J}_j-\sum_{i}{\bm h}\cdot\hat{\bm J}_i,
\label{eq:H}
\end{eqnarray}
where $\hat{\bm J}_i$ is the unit vector operator parallel or anti-parallel to the magnetic easy axis stemming from the CEF~\cite{WK2021,WPNAS} at the $i$th site 
and $J_{ij}$ is taken as the nearest-neighbor (N.N.) interaction $J_1$ and and the next-nearest-neighbor (N.N.N.) interaction $J_2$ for the intra IC and inter IC. 
The N.N. bonds and N.N.N. bonds for the intra IC and inter IC in the 1/1 AC are explained in Ref.~\cite{WI2024b} with illustration in Fig.~1. 
The second term in Eq.~(\ref{eq:H}) denotes the Zeeman term with the magnetic field ${\bm h}$. 

By performing the exact-diagonalization calculation, 
we numerically calculate the energies of all the states in the model (\ref{eq:H}) in the 1/1 AC with the 24 Tb sites inside the unit cell [see Fig.~\ref{fig:W_AW}(c)] under the periodic boundary condition and obtain the ground state with the lowest energy.  In this study, we analyze the whirling-anti-whirling state for $\theta=90^{\circ}$ shown in Fig.~\ref{fig:W_AW}(c).

We calculate the magnetization per site along the magnetic-field direction
\begin{eqnarray}
m=\frac{1}{N}\sum_{i}\hat{\bm J}_i\cdot\hat{\bm h}, 
\label{eq:mag}
\end{eqnarray}
where $\hat{\bm h}_i$ is the unit vector in the magnetic-field direction at the $i$th site. 

The topological charge $n$ on the IC was defined in Ref.~\cite{WPNAS} as 
\begin{eqnarray}
n=\frac{\Omega}{4\pi},
\label{eq:TC}
\end{eqnarray}
where $\Omega$ is the solid angle subtended by the 12 magnetic moments on the IC 
\begin{eqnarray}
\Omega=\sum_{i,j,k\in{\rm IC}}\Omega_{i,j,k}
\label{eq:SA}
\end{eqnarray}
and $4\pi$ is the surface area of the unit sphere. 
In Eq.~(\ref{eq:SA}), $\Omega_{i,j,k}$ is the solid angle subtended by the three magnetic moments at each triangular surface of the IC with $i$, $j$, and $k$ sites~\cite{Eriksson}
\begin{eqnarray}
\Omega_{ijk}=2\tan^{-1}\left[
\frac{\chi_{ijk}}{1+{\bm J}_i\cdot{\bm J}_j+{\bm J}_j\cdot{\bm J}_k+{\bm J}_k\cdot{\bm J}_i}
\right], 
\label{eq:OMGijk}
\end{eqnarray}
where $\chi_{ijk}$ is the scalar chirality defined as
\begin{eqnarray}
\chi_{ijk}={\bm J}_i\cdot({\bm J}_i\times{\bm J}_k)
\label{eq:sc}
\end{eqnarray}
and ${\bm J}_i$ is the magnetic moment vector at the $i$th site. 

We also calculate the total chirality 
\begin{eqnarray}
{\bm \chi}^{\rm T}=\sum_{\langle i,j,k\rangle}\chi_{ijk}\hat{n}_{ijk}
\label{eq:tot_c}
\end{eqnarray}
where $\hat{n}_{ijk}$ denotes the surface normal of the surface of the triangle $ijk$ sites on the IC and the summation is taken over the 20 triangle surfaces of the IC~\cite{WPNAS}. 

The total chirality plays as a role of the emergent fictious magnetic field, which causes the topological Hall effect in the electrical conductivity as~\cite{Taguchi2001,Tatara,Nagaosa,Tokura,Aoyama}
\begin{eqnarray}
\sigma_{\mu\nu}\propto\epsilon_{\mu\nu\rho}\chi^{\rm T}_{\rho},
\label{eq:THC}
\end{eqnarray}
where $\epsilon_{\mu\nu\rho}$ is the Levi-Civita symbol. 

We set $J_1=1.0$ as the energy unit throughout this paper.

\section{Results}

\subsection{Ground state at zero magnetic field}

Recently, the ground-state phase diagram of the effective model (\ref{eq:H}) in the 1/1 AC has been determined in the plane of $J_2/J_1$ and $\theta$ by the exact-diagonalization calculation~\cite{WI2024}. 
The strong FM interaction between the N.N.N. sites in the single IC stabilizes the whirling-moment state or anti-whirling moment state when the anisotropy angle $\theta$ is around $90^{\circ}$~\cite{WPNAS}. Then, the strong FM interaction between the N.N.N. sites for the inter IC  in the 1/1 AC stabilizes the whirling-anti-whirling state~\cite{WI2024}. 
Namely, the region where the whirling-anti-whirling state is stabilized in the phase diagram is around $\theta\approx 90^{\circ}$ for large $J_2/J_1$, whose feature was captured by the previous study~\cite{WI2024}.  
At $\theta=90^{\circ}$, the whirling-anti-whirling state [see Fig.~\ref{fig:W_AW}(c)] is realized for $J_2/J_1\ge 7.5$ with the FM interaction $J_1>0$~\cite{WPNAS}.  
Experimentally, the positive Curie-Weiss temperature was observed in the 1/1 AC Au-SM-Tb (SM=Al, Ga) where the whirling-anti-whirling order is realized~\cite{Sato2019,Nawa2023,Labib2024}. 
This implies that the interaction between the magnetic moments of Tb is FM. 
The model (\ref{eq:H}) with the FM interaction shows that the whirling-anti-whirling state is stabilized around $\theta=90^{\circ}$, which explains the experiments.
This suggests that the model (\ref{eq:H}) captures the essential feature of the magnetism in the 1/1 AC and therefore we consider (\ref{eq:H}) as the effective model. 

For the whirling-anti-whirling state, we have calculated the topological charge on the basis of Eq.~(\ref{eq:TC}). 
The obtained result shows that the whirling-moment state [see Figs.~\ref{fig:W_AW}(a), \ref{fig:W_AW}(b)] is characterized by the topological charge $n=3$ and anti-whirling-moment state is characterized by $n=-3$~\cite{WPNAS}. 
The total chirality is zero ${\bm \chi}^{\rm T}={\bf 0}$ in the whirling-moment state and the anti-whirling-moment state~\cite{WPNAS,WI2024}. 


\subsection{Effect of magnetic field}

The effect of the magnetic field for the whirling-anti-whirling state was analyzed theoretically in Ref.~\cite{WPNAS}. 
Recently, the ground-state property under the magnetic field applied along the 2-fold-axis direction has also been studied theoretically~\cite{WI2024}. 
Here, we analyze the effect of the magnetic field applied along the 3-fold-axis direction. 

\begin{figure}
\begin{center}
\includegraphics[width=10cm]{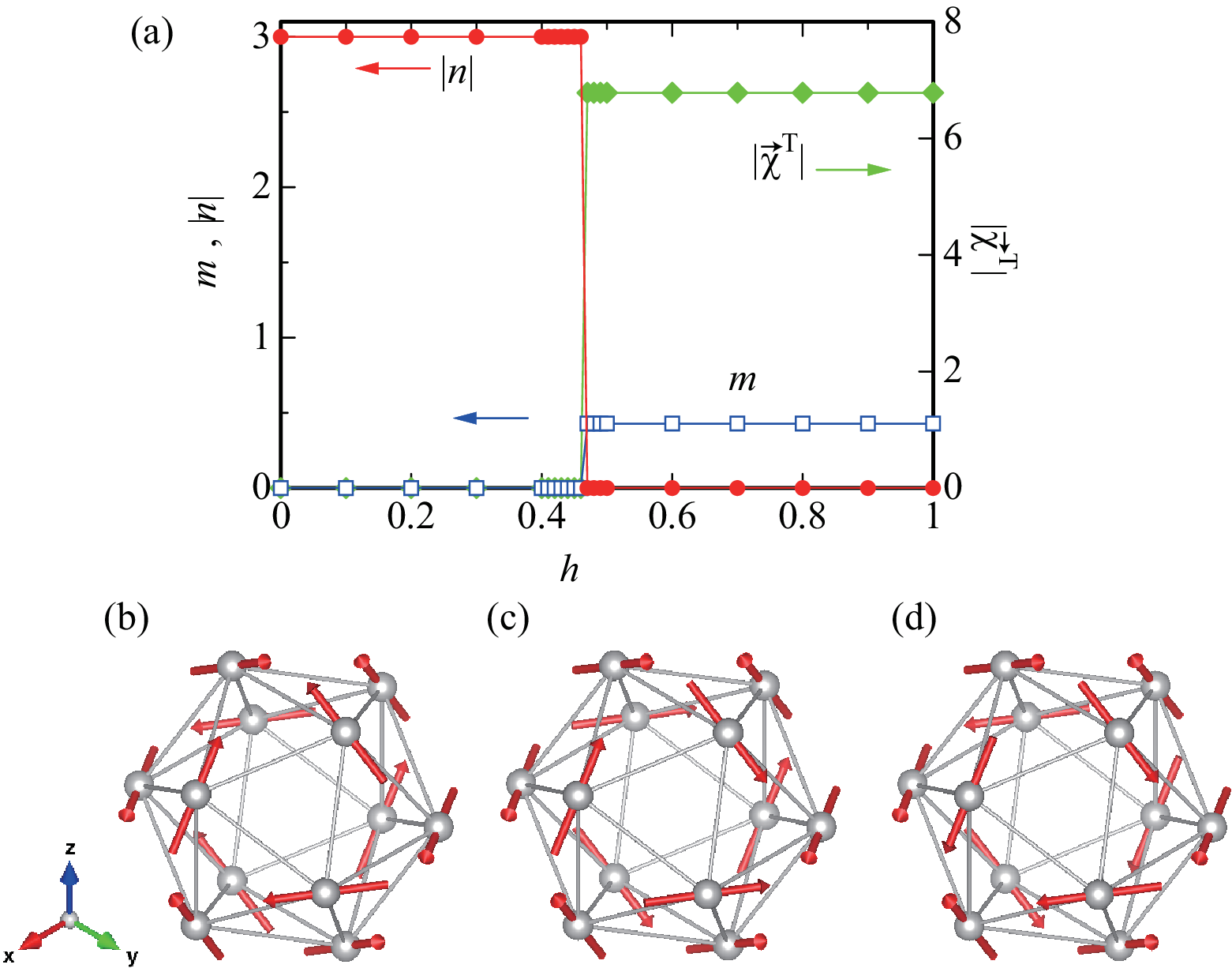}
\caption{\label{fig:m_h111}(a) Magnetic field dependence of the magnetization $m$ and the magnitude of the topological charge $|n|$ and the total chirality $|{\bm \chi}^{\rm T}|$. The magnetic field is applied along the (111) direction to the effective model~(\ref{eq:H}) with $J_1=1.0$ and $J_2=8.0$. 
(b), (c), (d) Magnetic states for $h>h_{\rm M}$ viewed from the (111) direction, which are energetically degenerate.  
}
\end{center}
\end{figure}

We have applied the magnetic field along the (111) direction to the whirling-anti-whirling state [see Fig.~\ref{fig:W_AW}(c)]. 
The result of the numerical calculation for $J_2=8.0$ at the anisotropy angle $\theta=90^{\circ}$ is shown in Fig.~\ref{fig:m_h111}(a). 
We find that the magnetization $m$ defined in Eq.~(\ref{eq:mag}) becomes finite from zero for $h\ge h_{\rm M}\equiv 0.47$. Namely, metamagnetic transition takes place at $h_{\rm M}$. 
For $h>h_{\rm M}$, the magnetic structure shown in Fig.~\ref{fig:m_h111}(b) is realized at both ICs located at the center and corner of the bcc unit cell.
This state is obtained by inverting the magnetic moments at the 6 sites in the whirling state and anti-whirling state shown in Fig.~\ref{fig:W_AW}(c) to earn the Zeeman energy.  

By calculating the topological charge defined in Eq.~(\ref{eq:TC}), this state shown in Fig.~\ref{fig:W_AW}(b) is turned out to be characterized by $n=0$. This implies that the topological transition takes place simultaneously with the metamagnetic transition at $h=h_{\rm M}$ as shown in Fig.~\ref{fig:m_h111}(a). 
We have also calculated the total chirality by Eq.~(\ref{eq:tot_c}) and have obtained the result as ${\bm \chi}^{\rm T}=(-5.64, 0.00, 3.76)$ for the magnetic state shown in Fig.~\ref{fig:m_h111}(b). 
This implies that the emergent fictious magnetic field due to the chirality of the noncoplanar magnetic moments appears after the metamagnetic transition for $h>h_{\rm M}$, as shown in Fig.~\ref{fig:m_h111}(a). 

We note here that the magnetic states shown in Figs.~\ref{fig:m_h111}(c), \ref{fig:m_h111}(d) are energetically degenerate into the state presented in Fig.~\ref{fig:m_h111}(b). We have confirmed that the results shown in Fig.~\ref{fig:m_h111}(a) are also obtained for these states. 
The total chirality is obtained as ${\bm \chi}^{\rm T}=(0.00,3.76,-5.64)$ for the state shown in Fig.~\ref{fig:m_h111}(c) and ${\bm \chi}^{\rm T}=(3.76, -5.64, 0.00)$ for the state shown in Fig.~\ref{fig:m_h111}(d).

\begin{figure}
\begin{center}
\includegraphics[width=14cm]{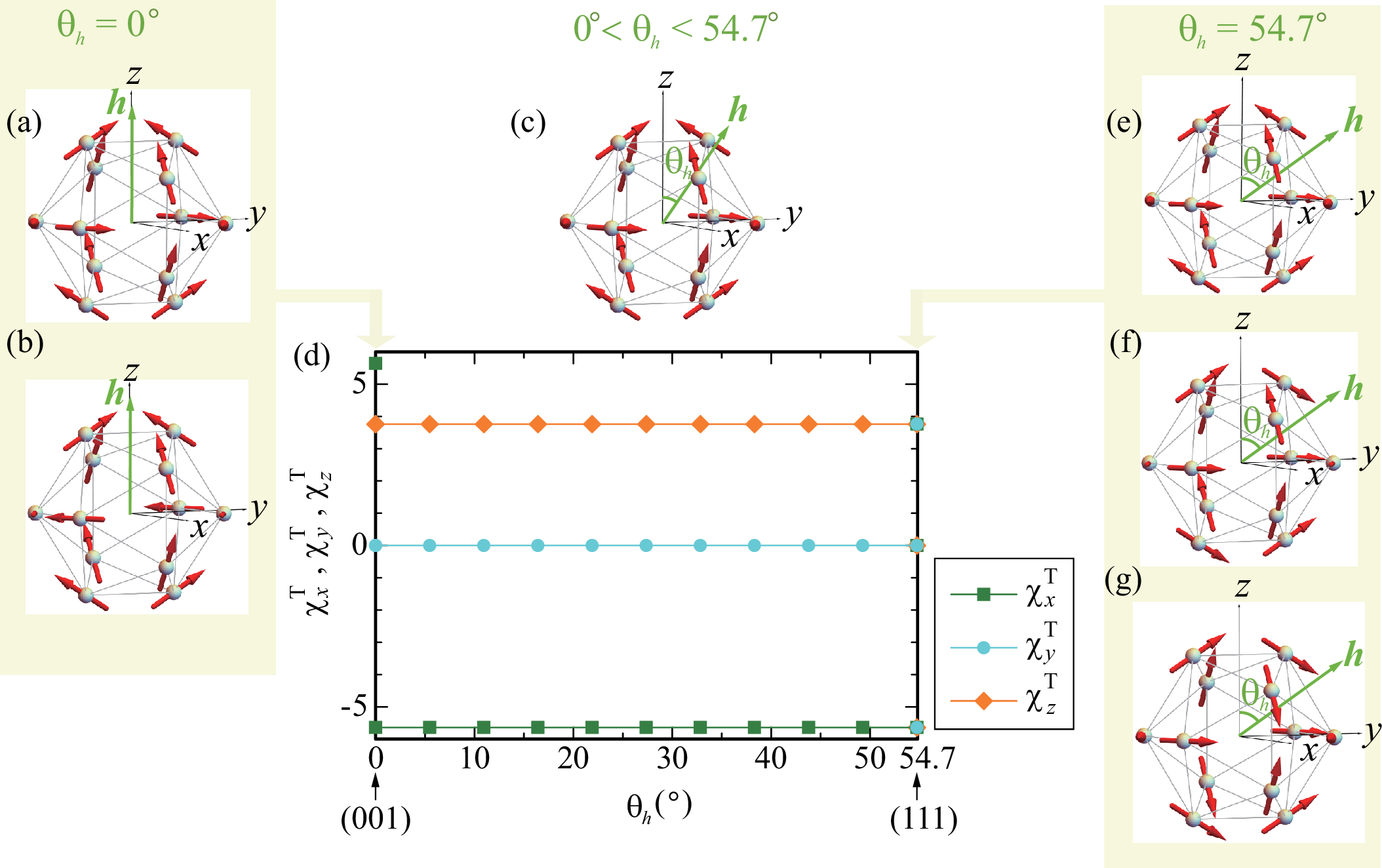}
\caption{
(a) The ground state at $J_1=1.0$, $J_2=8.0$, and ${\bm h}=(0.0,0.0,0.5)$. (b) Another ground state with the same energy as that of (a). (c) The ground state for $0^{\circ}<\theta_{\rm h}<54.7^{\circ}$.  
The angle $\theta_h$ is defined between the magnetic field ${\bm h}$ and the $z$ axis. 
(d) The field-direction dependence on the total chirality $\chi^{\rm T}_{x}$ (green square), $\chi^{\rm T}_{y}$ (light blue circle), and $\chi^{\rm T}_{z}$ (orange diamond) 
for $J_1=-1.0$, $J_2=-8.0$, and $|{\bm h}|=0.5$. 
The field direction is changed from the (001) direction to the (111) direction. 
(e), (f), (g) The ground states for ${\bm h}=(0.5,0.5,0.5)/\sqrt{3}$, which are energetically degenerate. 
}
\label{fig:tot_c_angle}
\end{center}
\end{figure}

Next, to clarify the field-direction dependence on the total chirality, we have changed the magnetic-field direction from ${\bm h}\parallel (001)$ to ${\bm h}\parallel (111)$ with keeping the magnitude $h=|{\bm h}|=0.5$. Here, the field direction is expressed by the angle $\theta_h$ defined between the magnetic field ${\bm h}$ and the $z$ axis.
For $\theta_h=0^{\circ}$, the magnetic field is applied along the $z$ axis, i.e., the 2-fold axis.  
In this case, the metamagnetic transition takes place at $h=h_{\rm M}'\equiv 0.43$~\cite{WI2024}. For $h>h_{\rm M}'$, the two magnetic states shown in Figs.~\ref{fig:tot_c_angle}(a), \ref{fig:tot_c_angle}(b) are energetically degenerate in the ground state, one of which shown in Fig.~\ref{fig:tot_c_angle}(a) is the same state as that presented in Fig.~\ref{fig:m_h111}(b). This state gives ${\bm \chi}^{\rm T}=(-5.64, 0.00, 3.76)$.  
In the another state, the total chirality is ${\bm \chi}^{\rm T}=(5.64,0.00,3.76)$, which is also plotted at $\theta_h=0^{\circ}$ in Fig.~\ref{fig:tot_c_angle}(d). 
Since the sign in $\chi^{\rm T}_x$ is opposite in these doubly degenerate states, the $x$ component of the total chirality is cancelled each other. 

When the magnetic field is deviated from the $z$ axis direction, namely for $\theta_h\ne0^{\circ}$, the degeneracy of the ground state is lifted reflecting the low symmetry of the field direction so that the unique ground state shown in Fig.~\ref{fig:tot_c_angle}(c) is realized.  This state is the same state presented in Fig.~\ref{fig:m_h111}(b). Hence, the total chirality ${\bm \chi}^{\rm T}=(-5.64,0.00,3.76)$ is realized for $0^{\circ}<\theta_h<54.7^{\circ}$. 

At $\theta_h=54.7^{\circ}$, the magnetic field is applied along the (111) direction, where the ground state for $h>h_{\rm M}$ is triply degenerate as presented in Figs.~\ref{fig:tot_c_angle}(e)-(g). The states shown in Figs.~\ref{fig:tot_c_angle}(e), \ref{fig:tot_c_angle}(f), \ref{fig:tot_c_angle}(g) are the same states as those presented  in Figs.~\ref{fig:m_h111}(b), \ref{fig:m_h111}(c), \ref{fig:m_h111}(d), respectively. The total chirality for the state shown in Fig.~\ref{fig:tot_c_angle}(f) is ${\bm \chi}^{\rm T}=(0.00,3.76,-5.64)$ and for the state shown in Fig.~\ref{fig:tot_c_angle}(g) is ${\bm \chi}^{\rm T}=(3.76,-5.64,0.00)$.  Hence, the three sets of ${\bm \chi}^{\rm T}$ are plotted at $\theta_h=54.7^{\circ}$ in Fig.~\ref{fig:tot_c_angle}(d). 

As for ${\bm h} \parallel (001)$ and ${\bm h} \parallel (111)$, the magnetic field is directed to the high symmetry point so that the doubly degenerate states and the triply degenerate states appear  respectively after the metamagnetic transition. This implies that in actual material, the domain structure appears even in the single crystal. In the case of ${\bm h} \parallel (001)$, the two degenerate states for $h>h_{\rm M}'$ results in the finite $z$ component of ${\bm \chi}^{\rm T}$ as analyzed above so that the topological Hall effect is expected to appear in the electrical conductivity $\sigma_{xy}$. In the case of ${\bm h} \parallel (111)$, $x$, $y$, and $z$ components are partly cancelled each other by the triply degenerate states for $h>h_{\rm M}$ so that the topological Hall effect is expected to appear in $\sigma_{xy}$, $\sigma_{yz}$, and $\sigma_{zx}$ by the remaining contributions of each component of ${\bm \chi}^{\rm T}$. 
As for the field direction between ${\bm h} \parallel (001)$ and ${\bm h} \parallel (111)$, the topological Hall effect is expected to appear in the electrical conductivity $\sigma_{xy}$ and $\sigma_{yz}$ by the emergent fictious magnetic field ${\bm \chi}^{\rm T}=(-5.64,0.00,3.76)$ for $0^{\circ}<\theta_h<54.7^{\circ}$.

\section{Summary}

We have discussed the effect of the magnetic field applied to the whirling-anti-whirling state along the (111) direction on the basis of the effective model (\ref{eq:H}). 
The field-induced metamagnetic transition is shown to take place simultaneously with the topological transition. After the metamagnetic transition, emergent fictious magnetic field due to the total chirality becomes finite, which gives rise to the topological Hall effect in the electrical conductivity. 
We have shown that the triply degenerate ground states appear after the metamagnetic transition reflecting the fact that the magnetic field is applied to high symmetry direction of the 3-fold axis. By changing the field direction between the (111) direction to the (001) direction, the degeneracy is lifted, which enables us to detect the topological Hall conductivity $\sigma_{xy}$ and $\sigma_{yz}$. 

The whirling-anti-whirling order and the metamagnetic behavior were actually observed in the 1/1 AC Au-SM-Tb (SM=Al, Ga)~\cite{Sato2019,Nawa2023,Labib2024}. Hence, it is interesting to detect the topological Hall effect predicted theoretically in this study, which is left for future measurements.

\section*{Acknowledgements}

The author greatly acknowledges K. Momma for providing the software VESTA~\cite{Momma2011} to draw the magnetic structure with kind instructions. 
This work was supported by JSPS KAKENHI Grant Numbers JP22H04597, JP22H01170, and JP23K17672.


\end{document}